\documentclass[pra,twocolumn,showpacs]{revtex4-1}
\usepackage{savesym}
\usepackage{amsmath}
\savesymbol{iint}
\usepackage{txfonts}
\restoresymbol{TXF}{iint}
\usepackage{amssymb}
\usepackage{mathrsfs}
\usepackage{bbm}
\usepackage{bm}
\usepackage{graphicx}

\newcommand{\ket}[1]{\left|#1\right\rangle}
\newcommand{\bra}[1]{\left\langle#1\right|}

\begin{document}

\title{Quantum coherence versus quantum discord in two coupled semiconductor double-dot molecules via a transmission line resonator}
\author{Pei Pei}
\email{ppei@mail.dlut.edu.cn}
\author{Chong Li}
\email{lichong@dlut.edu.cn}
\author{Jia-Sen Jin}
\author{He-Shan Song}
\email{hssong@dlut.edu.cn}
\affiliation{School of Physics and Optoelectronic Engineering,\\ Dalian University of Technology, Dalian 116024, People's Republic of China}
\date{November 9, 2010}
\begin{abstract}
We study the dynamics of quantum coherence and quantum correlations in two semiconductor double-dot molecules separated by a distance and indirectly coupled via a transmission line resonator. Dominant dissipation processes are considered. The numerical results show the sudden death of entanglement and the robustness of quantum discord to sudden death. Furthermore, the results indicate the dephasing processes in our model can lead in the revival and decay of coherence and discord with the absence of entanglement for certain initial states. By observing the dynamics of coherence versus discord for different initial states, we find that the similarities and differences of coherence and discord are not only related to the dependance of discord on optimizing the measurement set, but more importantly to the coherences in individual qubits which are captured by the adopted coherence measure.
\end{abstract}
 \maketitle

\section{Introduction}
Quantum entanglement, as one of the most intriguing hallmarks in the quantum world, manifests nonclassical correlation between quantum systems which cannot be found in any classical system \cite{Horodecki09}. Entanglement is a key ingredient for quantum communication, quantum cryptography, and quantum computing \cite{Nielsen00}. Besides the entanglement, other nonclassical correlations have been demonstrated \cite{Ollivier01,Henderson01,*Vedral03,Li07,*Luo08}. Ollivier and Zurek introduced the so called quantum discord (QD) \cite{Ollivier01} to describe the difference between two quantum expressions of the classical mutual information. Nonclassical correlation may still exist with the absence of entanglement, which implies the quantum discord as a more proper measure of the quantumness of correlations. When quantum discord is vanishing, the local accessibility of the classical information without perturbing the composite system is shown. Recently, quantum discord has received great attention on various topics, e.g., its potential to be the resource for deterministic quantum computation with one quantum bit \cite{Datta08,Lanyon08}, the immunity to the ``entanglement sudden death'' (ESD) phenomenon \cite{Werlang09,*Fanchini10,Ferraro10}, the sufficient condition for completely positive maps \cite{Rodriguez-Rosario08,Shabani09}, the QD close to the quantum phase transition \cite{Dillenschneider08,Sarandy09,Werlang10}, the behavior of QD in the Grover search \cite{Cui10}, the monogamic relation between the entanglement of formation (EoF) and the QD \cite{Fanchini10ar}, and the relation between entanglement irreversibility and the QD \cite{Cornelio10ar}.

On the other hand, among a number of physical systems \cite{Ladd10}, spins in semiconductor double-dot molecules (DDMs) are promising candidates for qubits to implement quantum information processing (QIP) in a solid-state system. Approaches based on DDMs combines spin and charge
manipulation and thus take advantage of the stability \cite{Taylor07}, the scalability \cite{Childress04}, and the efficiency for readout and coherent manipulation of charge states \cite{Petta05}. Several schemes adopting the architecture with DDMs have been proposed to show potential for low-noise coherent electrical control \cite{Taylor06}, to generate cluster states \cite{Lin08,*Lin09} and to implement quantum computing \cite{Xue10}. Moreover, Fanchini \textit{et al}. \cite{Fanchini10njp} investigate the dynamics of EoF and QD between two DDMs, considering the effects of dissipation. In their work, each DDM has one excess electron and the two DDMs are coupled through direct Coulomb interaction, which is fine for demonstrating the quantum correlations between two solid-state qubits. However, if the direct interaction between qubits is strong, it may be problematic when a third adjacent qubit is introduced. The direct interactions between the third qubit and the former two qubits may disturb the operation of the two-qubit gates and the control of nonadjacent qubits by indirect interaction may avoid this. This motivates us to  investigating the dynamics of quantum correlation between nonadjacent qubits, and verify whether the considered quantum correlations, QD and EoF, are resistant to dissipations in a scalable solid system. For this purpose, we adopt a quite different architecture: each DDM contains two excess electrons and the two DDMs are indirectly coupled through a transmission line resonator (TLR) \cite{Blais04,Wallraff04,Blais07} via a capacitor. Dissipation processes, e.g. the photon leakage, the pure dephasing and energy relaxation of qubits are taken into account for realistic.

Furthermore, we are interested in the relation of quantum correlations and quantum coherence (QC). Quantum correlation and coherence both arise from quantum superposition. If there exists coherence among subsystems of a compound quantum system, the nonlocal coherence may induce the nonclassical correlation between the corresponding subsystems, besides the local coherence of each corresponding subsystem \cite{Yu09}. In this case, quantum correlation is a kind of nonlocal coherence. We investigate the coherence dynamics from the point of off-diagonal elements of the density matrix describing the composite system. The effect of pure dephasing to the dynamics of QC and quantum correlations are presented and demonstrated. The similar and different behaviors of QC and QD are shown and discussed for varies initial states.

The paper is organized as follows. In Sec. II, we briefly review the basic concepts of quantum discord. In Sec. III, we describe the adopted architecture in detail and give the Hamiltonian of the system. The dominant dissipation processes, the typical system parameters, and the master equation are demonstrated. In Sec. IV, we present and discuss the numerical results for the effect of the pure dephasing and the behaviors of quantum coherence and quantum discord in two coupled DDMs. Finally a summary and some prospects are provided in section V.

\section{Quantum discord}
In classical information theory, the correlation between two systems $\mathcal{A}$ and $\mathcal{B}$ can be measured by the mutual information, $\mathcal{I}(\mathcal{A}:\mathcal{B})=\mathcal{H}(\mathcal{A})+\mathcal{H}(\mathcal{B})-\mathcal{H}(\mathcal{A},\mathcal{B})$, where $\mathcal{H}(\cdot)$ denotes the Shannon entropy $\mathcal{H}(p)=-\sum_{jk}p_{jk}\log_{2}p_{jk}$ \cite{Nielsen00} with the probability distribution $p$. The first extension of the expression to the quantum case is to directly replace the Shannon entropy by the von Neumann entropy $S(\rho_{\mathcal{A}\mathcal{B}})=-\textrm{Tr}(\rho_{\mathcal{A}\mathcal{B}}\log_{2}\rho_{\mathcal{A}\mathcal{B}})$, where $\rho_{\mathcal{A}\mathcal{B}}$ is the density operator describing the bipartite quantum system. The quantum mutual information between systems $\mathcal{A}$ and $\mathcal{B}$ is thus defined as
\begin{eqnarray}
\mathcal{I}(\rho_{\mathcal{A}\mathcal{B}})=S(\rho_\mathcal{A})+S(\rho_\mathcal{B})-S(\rho_{\mathcal{A}\mathcal{B}}),
\end{eqnarray}
where $\rho_\mathcal{A}=\textrm{Tr}_{\mathcal{B}}\rho_{\mathcal{A}\mathcal{B}}$, $\rho_\mathcal{B}=\textrm{Tr}_{\mathcal{A}}\rho_{\mathcal{A}\mathcal{B}}$ is the reduced density matrix obtained by taking the trace over all states of the system $\mathcal{B}$ or $\mathcal{A}$, respectively.

Through the Bayes rule, the classical mutual information can be equivalently rewritten as $\mathcal{J}(\mathcal{A}:\mathcal{B})=\mathcal{H}(\mathcal{A})-\mathcal{H}(\mathcal{A}|\mathcal{B})$, where $\mathcal{H}(\mathcal{A}|\mathcal{B})$ is the conditional entropy which quantifies the ignorance about the state of $\mathcal{A}$ on condition that the state of $\mathcal{B}$ is determined. When generalizing to the quantum case, the conditional entropy depends on what measurement is locally performed on $\mathcal{B}$ and what outcome is obtained. If one focuses on projective measurements described by a complete set of one-dimensional projectors $\{M_k\}$, the remaining state of $\mathcal{A}$ based on the outcome $k$ is $\rho_{\mathcal{A}|k}=(\mathbb{I}\otimes{M_k})\rho_{\mathcal{A}\mathcal{B}}(\mathbb{I}\otimes{M_k})/p_k$, where $\mathbb{I}$ is the identity operator performed on $\mathcal{A}$ and $p_k=\textrm{Tr}(\mathbb{I}\otimes{M_k})\rho_{\mathcal{A}\mathcal{B}}(\mathbb{I}\otimes{M_k})$ is the probability for obtaining $k$. The quantum conditional entropy is thus defined as the the residual entropy, $S(\rho_{\mathcal{A}}|\{M_k\})=\sum_{k}p_{k}S(\rho_{\mathcal{A}|k})$. The second extension of the classical mutual information to the quantum case can be written as the difference of entropy of $\mathcal{A}$ before and after the measurement on $\mathcal{B}$:
\begin{eqnarray}
\mathcal{J}(\rho_{\mathcal{A}\mathcal{B}})_{\{M_k\}}=S(\rho_{\mathcal{A}})-\sum_{k}p_{k}S(\rho_{\mathcal{A}|k}).
\end{eqnarray}
All the nonclassical correlations between $\mathcal{A}$ and $\mathcal{B}$ are removed by the local orthogonal projective measurement on $\mathcal{B}$ while contained in both the initial and residual entropies, therefore $\mathcal{J}(\rho_{\mathcal{A}\mathcal{B}})_{\{M_k\}}$ only captures the classical correlation between subsystems. $S(\rho_{\mathcal{A}}|\{M_k\})$ depends on the choice of $\{M_k\}$. To maximize $\mathcal{J}(\rho_{\mathcal{A}\mathcal{B}})_{\{M_k\}}$ one should optimize the set $\{M_k\}$ to seek the minimum of $S(\rho_{\mathcal{A}}|\{M_k\})$. Therefore the classical correlation is defined as \cite{Henderson01}
\begin{eqnarray}
\mathcal{C}(\rho_{\mathcal{A}\mathcal{B}})=S(\rho_{\mathcal{A}})-\textrm{min}_{\{M_k\}}\sum_{k}p_{k}S(\rho_{\mathcal{A}|k}).
\end{eqnarray}

The discrepancy between the two quantum extensions of classical mutual information defines the quantum discord \cite{Ollivier01}:
\begin{eqnarray}
\mathcal{Q}(\rho_{\mathcal{A}\mathcal{B}})=\mathcal{I}(\rho_{\mathcal{A}\mathcal{B}})-\mathcal{C}(\rho_{\mathcal{A}\mathcal{B}}).
\end{eqnarray}
For a bipartite quantum system, which is just the case in this paper, Hamieh \textit{et al}. \cite{Hamieh04} have demonstrated that the projective measurement is the positive operator valued measure which can maximize the classical correlations. This property is useful in our numerical calculation process.

\section{Description of the model}
The architecture we adopt consists of two identical DDMs capacitively coupled to a TLR, as shown in Fig. 1. Each DDM is formed by a GaAs/AlGaAs heterostructure with a layer of two-dimensional electron gas below. Several lithographically defined metallic gates are used to control the double-well potential for two dots and interdot tunneling \cite{Petta05}. With a modest external magnetic field $B_z=100\,\textrm{mT}$ along axis $z$, Zeeman effect results in energy gaps between the spin-aligned states $\ket{(1,1)T_+}=\ket{\upuparrows}$, $\ket{(1,1)T_-}=\ket{\downdownarrows}$ and the spin-anti-aligned states $\ket{(1,1)T_0}=(\ket{\uparrow\downarrow}+\ket{\downarrow\uparrow})/\sqrt{2}$, $\ket{(1,1)S}=(\ket{\uparrow\downarrow}-\ket{\downarrow\uparrow})/\sqrt{2}$. The notation $(n_l,n_r)$ labels $n_l$ electrons localized in the left dot and $n_r$ electrons in the right dot. Follow the approximation of Ref.~\cite{Lin08,Xue10,Guo08}, each DDM can be deduced to a two-level system in the basis $\{\ket{(1,1)S},\ket{(0,2)S}\}$, described by the Hamiltonian
\begin{eqnarray}
H_d=-\Delta\ket{(0,2)S}\bra{(0,2)S}+T\left(\ket{(1,1)S}\bra{(0,2)S}+\textrm{h.c.}\right),
\end{eqnarray}
where $\hbar=1$, and $\Delta$ and $T$ are the relative potential and tunneling \cite{Taylor07} between $\ket{(0,2)S}$ and $\ket{(1,1)S}$, respectively. The energy gap between the eigenstates is $\omega_d=\sqrt{4T^2+\Delta^2}$. Here we choose the optimal point $\Delta=0$ for maximum coupling \cite{Xue10}, and use the resulting $\ket{1}=(\ket{(0,2)S}+\ket{(1,1)S})/\sqrt{2}$ and $\ket{0}=(\ket{(0,2)S}-\ket{(1,1)S})/\sqrt{2}$ as qubit states.

\begin{figure}
\includegraphics[width=8cm]{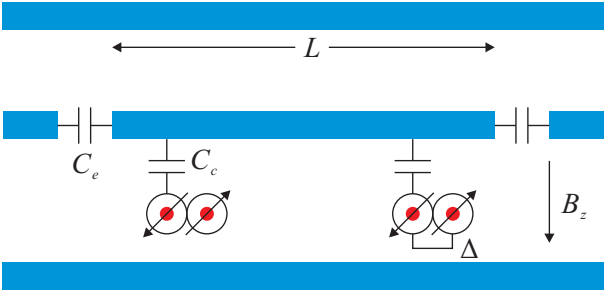}
\caption{\label{fig:1}(Color online) Schematic diagram for the adopted architecture. The TLR has a length of $L$, the DDMs has a relative potential $\Delta$ between two singlet states. The two DDMs are individually coupled two the TLR via the identical capacitor $C_c$. The TLR is connected to the input/output ports via a capacitor $C_e$ for writing/reading signals.}
\end{figure}

We consider the TLR with length $L$, capacitance per unit length $C_0$, and characteristic impedance $Z_0$. The DDM is on resonance with the fundamental frequency of TLR, $\omega_0=\pi/LC_{0}Z_0$, therefore we can neglect all the higher energy modes of the TLR \cite{Childress04}. The Hamiltonian of the TLR reads
\begin{eqnarray}
H_r=\omega_{0}a^{\dag}a,
\end{eqnarray}
where $a$ $(a^{\dag})$ is the annihilation (creation) operator of the resonator.

As shown in Fig. 1, the two DDMs are individually coupled to the TLR via the identical capacitor $C_c$. The charging energy provides the DDM-TLR interaction, $H_{int}=C_cV_rV_d$, where $V_r$ and $V_d$ are the voltages of the TLR and DDM, respectively. The Coulomb interaction between the two DDMs can be neglected due to the long distance ($\sim4\mu\textrm{m}$) between them with respect to the small size of DDM ($\sim400\textrm{nm}$). After the quantization of the voltages of the TLR \cite{Blais04}, $V_r=\sqrt{\omega/(LC_0)}(a^{\dag}+a)$ and the DDM \cite{Guo08}, $V_d=2e/C_{tot}\ket{(0,2)S}\bra{(0,2)S}+e/C_{tot}\ket{(1,1)S}\bra{(1,1)S}$, and in the rotating wave approximation, the full Hamiltonian takes the usual Jaynes-Cummings form
\begin{eqnarray}
H=\frac{1}{2}\sum_{i=1,2}\omega_{di}\sigma_{zi}+\omega_{0}a^{\dag}a+g\sum_{i=1,2}(a\sigma_{i}^{\dag}+\textrm{h.c.}),
\end{eqnarray}
where $\sigma_z$ is the pauli $z$ operator, $\sigma^{\dag}=\ket{1}\bra{0}$, $\sigma=\ket{0}\bra{1}$, and the coupling coefficient
\begin{eqnarray}
g=\frac{eC_c}{2C_{tot}}\sqrt{\frac{\omega_0}{LC_0}}.
\end{eqnarray}
Here $C_{tot}$ is the total capacitance of the DDM. As mentioned above, we consider the two DDMs identical and resonant with the fundamental frequency of TLR, $\omega_{d1}=\omega_{d2}=\omega_0$, thus the Hamiltonian in the interaction picture reads
\begin{eqnarray}
V=g\sum_{i=1,2}(a\sigma_{i}^{\dag}+\textrm{h.c.}).
\end{eqnarray}

Now we demonstrate the dominant dissipation processes considering the interaction with environment in the Born-Markov approximation. First, the photon leakage of the resonator which can be described by the damping rate $\kappa=\omega_{0}/Q$, where $Q$ is the quality factor of the TLR. Though the internal loss of the resonator is negligible ($Q_{int}$ up to $10^6$) \cite{Blais04,Wallraff04}, with the external magnetic field $B_z$ the quality factor will be reduced to $Q\sim10^4$ \cite{Frunzio05}. The TLR frequency chosen as $\omega_{0}/2\pi=10\,\textrm{Ghz}$ leads to the damping rate $\kappa/2\pi=\omega_{0}/2\pi{Q}=1\,\textrm{Mhz}$. The second is the spin relaxation induced by the qubits coupling to the phonon bath. The induced spin relaxation time is obatined as $T_1\sim1\,\mu\textrm{s}$ using the spin-boson model at the optimal point \cite{Taylor06}. The third is the spin dephasing results from the low frequency fluctuations of
the electrostatic bias and the hyperfine interaction with the nuclear spins. The former dephasing time is estimated as $T_2\sim\omega_dT_b^2\approx10\,\textrm{ns}$ \cite{Taylor06}, with the observed bare dephasing time $T_b\approx1\,\textrm{ns}$ \cite{Petta04}. The latter time-ensemble-averaged dephasing time is measured as $T_2^\ast\sim10\,\textrm{ns}$ and by spin-echo technique the dephasing time may be prolonged beyond $1\,\mu\textrm{s}$ \cite{Petta05}. Taking into account the above processes, the evolution of the state of the three party system (the two DDMs plus the TLR) can be described by the master equation \cite{Blais07}
\begin{widetext}
\begin{eqnarray}
\dot{\rho}=-i[V,\rho]+\sum_{i=1,2}\frac{\gamma_{\varphi,i}}{2}(\sigma_{zi}\rho\sigma_{zi}-\rho)+
\sum_{i=1,2}\frac{\gamma_i}{2}(2\sigma_{i}\rho\sigma_{i}^{\dag}-\sigma_{i}^{\dag}\sigma_{i}\rho-\rho\sigma_{i}^{\dag}\sigma_{i})+
\frac{\kappa}{2}(2a\rho{a}^{\dag}-a^{\dag}a\rho-\rho{a^{\dag}}a),\label{me}
\end{eqnarray}
\end{widetext}
where $\gamma_{\varphi,i}$ and $\gamma_i$ are the pure dephasing rate and relaxation rate, respectively. We choose $\gamma_{\varphi,1}=\gamma_{\varphi,2}=\gamma_{\varphi}$ and $\gamma_1=\gamma_2=\gamma$ for identical DDMs. In simulation the coupling coefficient is chosen as $g/2\pi=100\,\textrm{Mhz}$ which has been experimentally realized \cite{Wallraff04}. To show the effect of pure dephasing clearly we do not assume the spin-echo technique is applied; therefore with the spin relaxation and dephasing times demonstrated above, the damping rates are translated to $\gamma_{\varphi}/2\pi=15.8\,\textrm{Mhz}$ and $\gamma/2\pi=0.16\,\textrm{Mhz}$. In simulation the photon number of the resonator is truncated at the value of $5$.

\section{Quantum coherence versus quantum discord}
Before present the results it is necessary to describe the quantification of quantum coherence and the optimization of the measurement to obtain the quantum discord. Through tracing over the states of the resonator, we get the reduced density operator $\rho_{\mathcal{AB}}=\textrm{Tr}_{r}\rho$, which describes the state of the composite system consisting of the two DDMs, and $\mathcal{A}$, $\mathcal{B}$ labels the individual DDM, respectively. To quantify the coherence of the composite system, we focus on the the off-diagonal elements of $\rho_{\mathcal{AB}}$ in the basis $\{\ket{11}_{\mathcal{AB}},\ket{10}_{\mathcal{AB}},\ket{01}_{\mathcal{AB}},\ket{00}_{\mathcal{AB}}\}$. Yu and Song \cite{Yu09} have demonstrated the coherence in given basis can be quantified by the distance between the quantum state $\rho_{\mathcal{AB}}$ and the nearest incoherent state. Thus the coherence of $\rho_{\mathcal{AB}}$ can be measured by
\begin{eqnarray}
\mathcal{D}(\rho_{\mathcal{AB}})=\|\rho_{\mathcal{AB}}-\sigma^\ast\|_1=\sum_{i\neq{j}}|{\rho_{\mathcal{AB}_{ij}}}|,
\end{eqnarray}
where $\|\cdot\|_1$ is the ``entrywise'' norm and $\sigma^\ast$ is the diagonal matrix with the identical diagonal elements of $\rho_{\mathcal{AB}}$. We note the measure $\mathcal{D}$ includes the contribution of all off-diagonal elements of $\rho_{\mathcal{AB}}$, therefore it also contains the coherence of individual DDM $\mathcal{A}$ and $\mathcal{B}$, which will have influence on the numerical results of the coherence dynamic for certain states. In the calculation of quantum discord, we choose the measurement set on $\mathcal{B}$ as the projectors $\{\ket{\psi_1}\bra{\psi_1},\ket{\psi_2}\bra{\psi_2}\}$, in which $\ket{\psi_1}=\cos\theta\ket{0}+e^{i\phi}\sin\theta\ket{1}$ and $\ket{\psi_2}=\sin\theta\ket{0}-e^{i\phi}\cos\theta\ket{1}$. The parameters $\theta$ and $\phi$ are numerically varied from $0$ to $2\pi$ in order to seek the maximum of classical correlation $\mathcal{C}$ \cite{Werlang09}.

\begin{figure}
{\includegraphics[width=8cm]{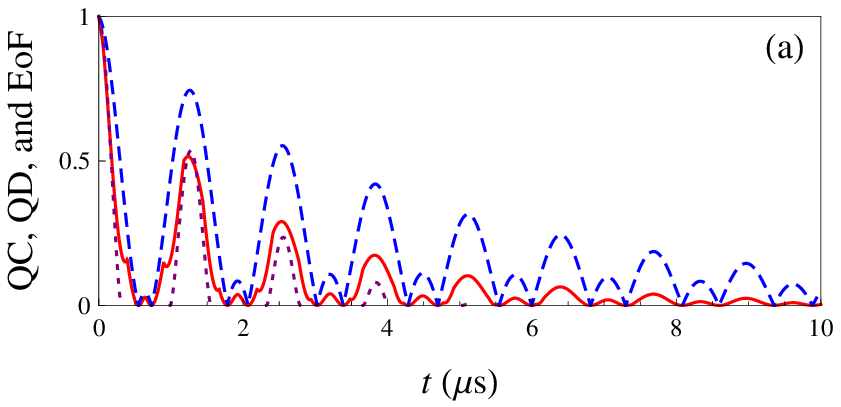}}
{\includegraphics[width=8cm]{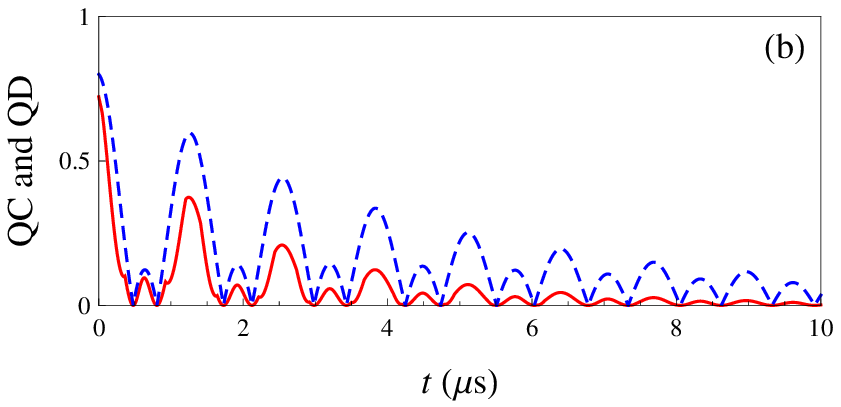}}
{\includegraphics[width=8cm]{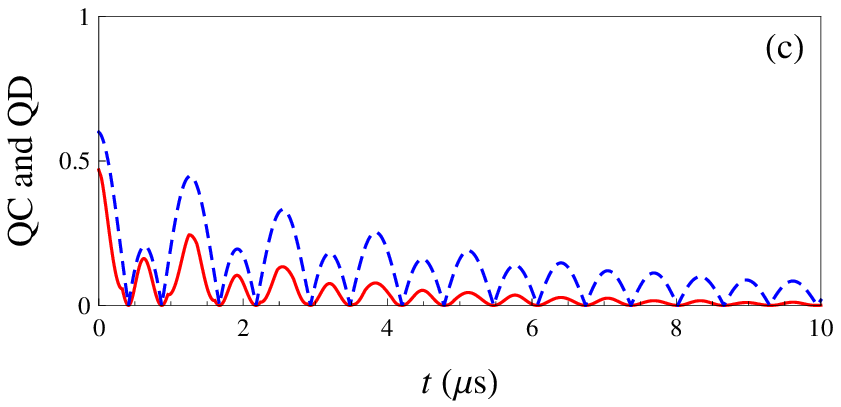}}
\caption{\label{fig:2}(Color online) Dynamics of quantum coherence and quantum correlations as a function of time. The solid (red), dashed (blue), and dotted (purple) curves correspond to the dynamics of discord, coherence, and EoF, respectively. The two DDM qubits are initially prepared in the state given in Eq. (\ref{belllikepsi}) with (a) $\alpha^2=1/2$, (b) $\alpha^2=1/5$, and (c) $\alpha^2=1/10$.}
\end{figure}

By numerical methods we obtain the dynamics of QC and QD in the two DDMs for different initial states. Firstly, we consider one kind of Bell-like state
\begin{eqnarray}
\ket{\psi_i}_\mathcal{AB}=\alpha\ket{01}_\mathcal{AB}+\sqrt{1-\alpha^2}\ket{10}_\mathcal{AB},\label{belllikepsi}
\end{eqnarray}
while the initial photon number in the TLR is assumed to be nonzero. For comparison we also calculate the entanglement dynamics for the case $\alpha^2=1/2$, using the EoF as the measurement of entanglement. For two qubits the EoF can be written in terms of Wootters' concurrence \cite{Wootters98}, $E(t)=-f(t)\log_2{f(t)}-(1-f(t))\log_2{(1-f(t))}$, where $f(t)=(1+\sqrt{1-\tilde{C}(t)^2})/2$. The concurrence is expressed as $\tilde{C}(t)=\textrm{max}\{0,\lambda_1-\lambda_2-\lambda_3-\lambda_4\}$, where $\lambda_i\,(i=1,2,3,4)$ are the square roots of the eigenvalues of $\rho_\mathcal{AB}\sigma_y\otimes\sigma_y\rho_\mathcal{AB}^\ast\sigma_y\otimes\sigma_y$ in decreasing order. The results are shown in Fig.~\ref{fig:2}(a)-\ref{fig:2}(c). We observe that the QC, QD, and EoF all present oscillations and decays. For QC and QD there are two different types of amplitude of oscillation. The first type of amplitude undergoes decays after respective revivals, while the second type of amplitude increases and then decreases for $\alpha^2=1$ and only decays for other values of $\alpha$ (an empty resonator also leads in this type of amplitude). For any pure state the QD and EoF coincide and for the maximally entangled state $\mathcal{D}=1$, thus for $\alpha^2=1/2$ the QC, QD, and EoF coincide initially. For non-maximally entanglement states the QC and QD certainly have different initial values. The EoF manifests the ESD and ``sudden birth'' phenomenons, which are the natural consequences under the combination action of the dissipation processes and the state transfer between DDMs and photons via the coupling to the TLR. Within short time intervals the EoF permanently vanishes (see Fig.~\ref{fig:2}(a)) which is much earlier than the QC and QD, indicating the environment has greater impact on the EoF than QC and QD. Furthermore, the QD and QC vanish only at discrete instants and manifest the second type of amplitude of oscillation within the vanishing intervals of the EoF dynamics. This interesting feature shows some specific interaction with the environment does not induce the regular decay and revival of nonclassical correlations \cite{Fanchini10} (up to one vanishing instant), but leads to the revival of the nonclassical correlation as well as the coherence to a maximum then decays (two vanishing instants). In our model this feature is induced by the interaction with nuclear spins and low frequency fluctuations of the electrostatic bias, which are characterized by the pure dephasing term in the master equation in Eq.~(\ref{me}). The other feature is that the QC and QD exhibit homology to a considerable degree throughout for Bell-like states $\ket{\psi_i}_\mathcal{AB}$. The QC and QD even vanish simultaneously. The discrepancy of the behaviors like the irregular shape of QD results from the dependance of QD on optimizing the measurement while QC is independent of measurement. We note that for Bell-like states $\rho_{\mathcal{AB}}$ has the $X$ structure and the coherences of individual DDMs are always zero. Thus the QC for this case only captures the coherence of the joint two-qubit system. The results indicate that if there is no coherence in subsystems, the coherence dynamic of the composite system shows the similar behavior of quantum discord, even with the absence of entanglement.

\begin{figure}
{\includegraphics[width=8cm]{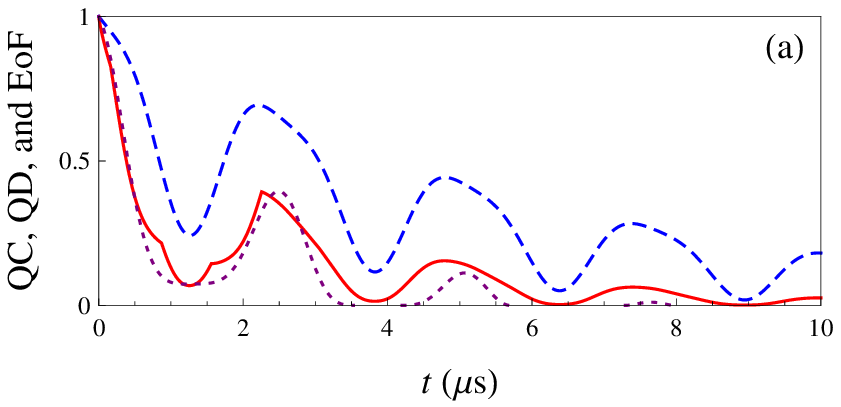}}
{\includegraphics[width=8cm]{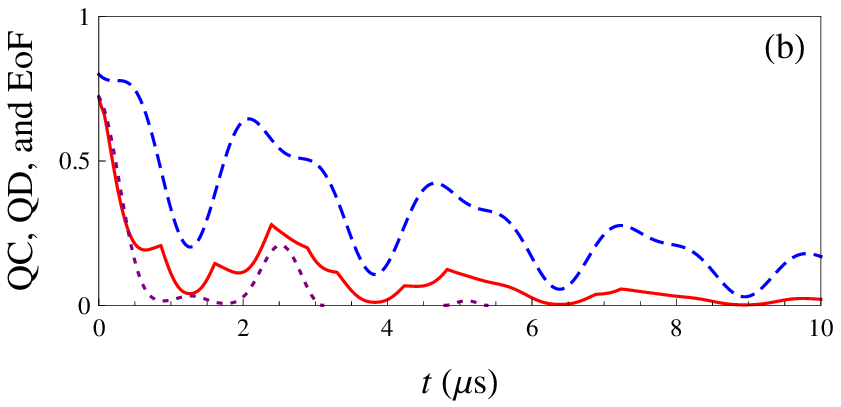}}
{\includegraphics[width=8cm]{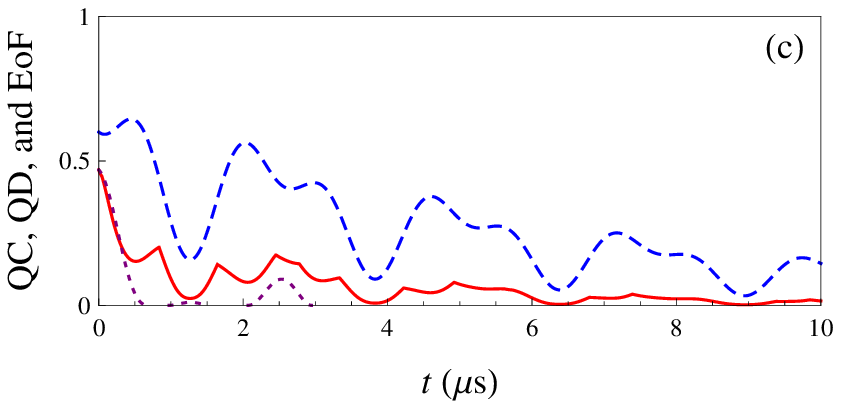}}
\caption{\label{fig:3}(Color online) Dynamics of quantum coherence and quantum correlations as a function of time. The solid (red), dashed (blue), and dotted (purple) curves correspond to the dynamics of discord, coherence, and EoF, respectively. The two DDM qubits are initially prepared in the state given in Eq. (\ref{belllikephi}) with (a) $\beta^2=1/2$, (b) $\beta^2=1/5$, and (c) $\beta^2=1/10$.}
\end{figure}

Next we consider the initial states as the other kind of Bell-like state
\begin{eqnarray}
\ket{\phi_i}_\mathcal{AB}=\beta\ket{00}_\mathcal{AB}+\sqrt{1-\beta^2}\ket{11}_\mathcal{AB},\label{belllikephi}
\end{eqnarray}
while the TLR is assumed to be initially empty. The results are shown in Fig.~\ref{fig:3}(a)-\ref{fig:3}(c). The behaviors of the QC, QD, and EoF are very different from those with the initial state $\ket{\psi_i}_\mathcal{AB}$. The three dynamics present the similar behavior for $\beta^2=1/2$, but for $\beta^2<1/2$ the oscillations and decays are quite complicated and exhibit respective trends. For some instants the QC and QD achieve a maximum while the EoF comes to a minimum, however despite the discrepancy of behavior around the sudden changes of QD, the dynamics of QC and QD are similar, especially after a relative long time. The QD still manifests the robustness to the ``sudden death'' which is in agreement with previous works. Besides, unlike the first case the QC does not vanish within finite time intervals.

\begin{figure}
\includegraphics[width=8cm]{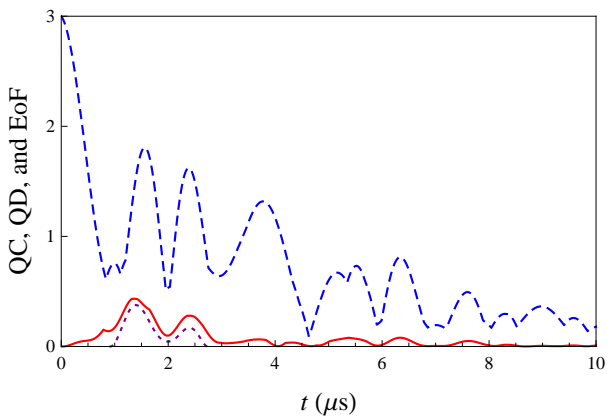}
\caption{\label{fig:4}(Color online) Dynamics of quantum coherence and quantum correlations as a function of time. The solid (red), dashed (blue), and dotted (purple) curves correspond to the dynamics of discord, coherence, and EoF, respectively. The two DDM qubits are initially prepared in the separable state given in Eq. (\ref{separable}).}
\end{figure}

At last we plot the dynamics of QC, QD, and EoF as a function of time with the two DDMs prepared in a specific separable state
\begin{eqnarray}
\ket{\varphi_i}_\mathcal{AB}=\frac{1}{2}(\ket{0}_\mathcal{A}+\ket{1}_\mathcal{A})\otimes(\ket{0}_\mathcal{B}+\ket{1}_\mathcal{B}),\label{separable}
\end{eqnarray}
as shown in Fig.~\ref{fig:4}. Through the interaction with the photons in the resonator, the entanglement between the two DDMs is generated then oscillates until vanishing. Coupling to an empty resonator will lead to no entanglement but still the generation of nonclassical correlations, in accordance with the results for the case of mediating by a common reservoir in Ref.~\cite{Fanchini10}. One notices that the behaviors of QC and QD differ widely, especially at the beginning, presenting substantial discrepancy between the initial separable state $\ket{\varphi_i}_\mathcal{AB}$ and the Bell-like states; however after a relative long time, the behaviors of QC and QD become similar. This is because for the initial state $\ket{\varphi_i}_\mathcal{AB}$, there is no nonclassical correlations between the two DDMs but exist coherences in individual qubits, which are nonzero within short time intervals. In this case, the QC measure $\mathcal{D}$ captures not only the coherence of the composite system, but also the coherences in subsystems, while the QD measure $\mathcal{Q}$ only quantifies the nonclassical correlations between subsystems. After a long time the cohereces in individual DDMs have been permanently lost due to the dissipations and the dynamics resembles the Bell-like cases, therefore the behaviors of QC and QD become similar afterwards.

In experiments, the measurement of the entanglement between distant DDMs is possible by the dispersive quantum nondemolition (QND) readout of qubits. In the dispersive regime, $g/(\omega_d-\omega_0)\ll1$, the QND readout of qubits can be realized by microwave irradiation of the resonator and then probing the transmitted or reflected photons. For example, by driving the resonator with a microwave pulse centered at the pulled frequency of $\omega_0+g^2/(\omega_d-\omega_0)$, the information about the state of the qubit is mostly stored in the number of transmitted photons. The scheme can be extended to the case of two or multiple qubits with different detunings (different DDMs) \cite{Blais04}, and shows the feasibility of a joint measurement of entanglement in our system.

\section{Summary and prospect}
In summary, we have numerically investigated the dynamics of quantum coherence and quantum correlations between two semiconductor DDMs separated by a distance and indirectly coupled via a transmission line resonator. The stronger impact on the EoF than QC and QD is verified, as well as the ESD phenomenon of EoF and the robustness of the QD to the sudden death, which implies the QD as a promising quantum information resource to process quantum computation in large scale solid systems. Furthermore, we find that the dephasing processes in our model induce a revival and decay of QC and QD within the vanishing interval of EoF for certain Bell-like states. We show that for initial Bell-like states, the QC and QD present similar behavior to a certain degree even with the absence of the entanglement, but for a initial specific separable state the behaviors are totally different at first but become similar after a long time. We find that the local coherences in individual qubits play an important role to the similarities and differences of the dynamics of QC and QD. The specific internal relation between coherence and discord of bipartite mixed states may be revealed in an exactly solvable model while considering a more proper coherence measure, e.g. some measure connected with the localizable coherence \cite{Yu09}. This inspires us to follow up in further research.

\begin{acknowledgements}
We thank Drs. Chang-Shui Yu and Song-Lin Wu for helpful discussions. This work is supported by National Natural Science Foundation of China (NNSFC) under Grants No. 10875020 and No. 60703100, and the Fundamental Research Funds for the Central Universities under Grant No. DUT10LK10.\\
\end{acknowledgements}

\end{document}